\documentclass[a4paper,11pt]{article}
\usepackage{pos}

\usepackage{lineno}

\usepackage{hyperref}
\usepackage[nooneline]{subfigure}
\subfiguretopcaptrue

\title{Evaluation of the effective mirror area of CTA Small-Sized Telescopes for camera design and Monte Carlo simulation}
 \ShortTitle{Effective Mirror Area of CTA SST}

\author*[a,b,c]{Akira Okumura}
\author[d]{Duncan Ross}
\author[e,f]{Francesco G. Saturni}
\author[g]{Giorgia  Sironi}
\author[h]{Richard  White}

\affiliation[a]{Institute for Space--Earth Environmental Research, Nagoya University,\\Furo-cho, Chikusa-ku, Nagoya 464-8601, Japan}
\affiliation[b]{Kobayashi--Maskawa Institute for the Origin of Particles and the Universe, Nagoya University,\\Furo-cho, Chikusa-ku, Nagoya 464-8602, Japan}
\affiliation[c]{Nagoya University Southern Observatories, Nagoya University,\\Furo-cho, Chikusa-ku, Nagoya 464-8602, Japan}
\affiliation[d]{University of Leicester, Space Research Centre, Space Park Leicester LE4~5SP, United Kingdom}
\affiliation[e]{INAF - Astronomical Observatory of Rome, Via Frascati 33, I-00078 Monte Porzio Catone (RM), Italy}
\affiliation[f]{ASI - Space Science Data Center, Via del Politecnico snc, I-00133 Roma, Italy}
\affiliation[g]{INAF - Astronomical Observatory of Brera, Via E. Bianchi, 46, 23807 Merate LC, Italy}
\affiliation[h]{Max-Planck-Institut für Kernphysik, P.O. Box 103980, D69029 Heidelberg, Germany}

\emailAdd{oxon@mac.com}

\abstract{The effective mirror area of an imaging atmospheric Cherenkov telescope is a crucial key parameter for trigger threshold determination and energy calibration. It is usually calculated by 3D ray-tracing simulation using a simplified telescope model, and the result is used in Monte Carlo simulations. However, simplified telescope and camera models are not adequate for the Schwarzschild–Couder configuration to be used in Small-Sized Telescopes (SSTs) of the Cherenkov Telescope Array. This is because the complex 3D structure of the secondary mirror, telescope masts, and camera body block a significant fraction of Cherenkov and night-sky photons. To evaluate the effective mirror area of an SST and to finalize its camera body design with minimal shadowing, a complex 3D model was built and simulated using the ROBAST ray-tracing library. A camera body size of 570\,mm and a window size of 430\,mm were selected for the final camera design based on the evaluation of shadowing by simulation. A non-axisymmetric effective area distribution was determined via the modeling of the complex telescope structure, while meeting the SST effective area requirement.}

\ConferenceLogo{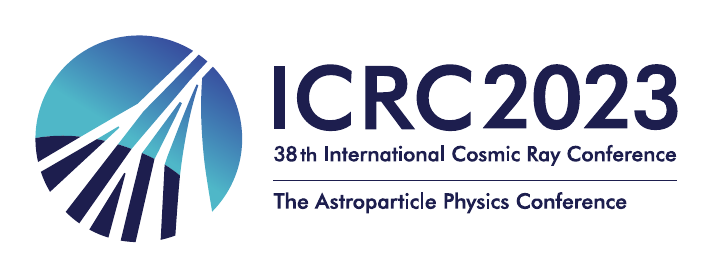}

\FullConference{%
The 38th International Cosmic Ray Conference (ICRC2023)\\
  26 July -- 3 August, 2023\\
  Nagoya, Japan}

\begin{document}
\maketitle

\section{Introduction}

The effective mirror area of gamma-ray and cosmic-ray telescopes is a key parameter used in Monte Carlo (MC) simulations, in which the photon tracks of atmospheric Cherenkov or fluorescence radiation are traced. In addition to understanding the optical properties of the atmosphere and focal-pane photodetectors, an accurate evaluation of the effective mirror area can improve the energy calibration and MC performance studies of the telescopes.

The optical systems conventionally used for gamma-ray and cosmic-ray observations are parabolic and Davies--Cotton (DC) systems. The evaluation of the effective area of these systems is not technically difficult or complex because they are (segmented) single-mirror telescopes. In these systems, the specular reflection of photons occurs only once on the mirror surface. Therefore the obscuration by the telescope structure (``shadowing'') is considered only before the reflection in most cases.

In addition to the conventional telescope designs, the Schwarzschild--Couder (SC) configuration has been proposed for future imaging atmospheric Cherenkov telescopes (IACTs) \cite{Vassiliev:2007:Wide-field-aplanatic-two-mirror-telescopes-for-gro}, which have aspherical primary and secondary mirrors. This configuration enables us to simultaneously realize a wide field of view (FOV) and high angular resolution using a compact small-plate-scale camera. However, calculating the effective area of an SC system is complex because the telescope masts supporting the secondary mirror and camera are located between the primary and secondary mirrors. Therefore, incident photons may be obscured by the masts before being reflected by the primary mirror and again before being reflected by the secondary mirror. In addition, the camera may block photons if its body diameter is excessively large. Therefore, non-sequential ray-tracing simulation and accurate three-dimensional (3D) modeling of the telescope and camera structure are required to evaluate the effective mirror area, which has been approximated in previous simplified simulations.

The use of the SC configuration in IACTs has been realized in prototype telescopes of the Cherenkov Telescope Array (CTA) \cite{Adams:2021:Detection-of-the-Crab-Nebula-with-the-9.7-m-protot,Lombardi:2020:First-detection-of-the-Crab-Nebula-at-TeV-energies}. CTA is the next-generation ground-based gamma-ray observatory that will be comprised of different telescope designs: Large-Sized Telescopes (LSTs, segmented 23\,m parabola), Medium-Sized Telescopes (MSTs, 12\,m DC and 10\,m SC), and Small-Sized Telescopes (SSTs, 4\,m SC). Among the four different telescope designs of CTA, this paper focuses on the effective mirror area evaluation of the SC optical system of the SSTs.

To meet the CTA requirement for the minimum effective area of the SSTs and to finalize the camera body design, we first calculate the effective mirror area as a function of different camera body sizes in Section~\ref{sec:camera_design}. In Section~\ref{sec:final_design}, we calculate the effective area with the final camera design, including its various components that can potentially block photon tracks between the primary and secondary mirrors. The nonuniformity of the effective area in the camera FOV is discussed as well.

\section{SST Camera Design}
\label{sec:camera_design}

CTA SST requirements include a camera FOV radius larger than $4^\circ$ (namely $8^\circ$ diameter) and an effective mirror area larger than 5\,$\mathrm{m}^2$ over the FOV, without considering the losses caused by photodetector gaps and camera window transmittance. To meet these requirements, not only the telescope structure but also the camera body size must be carefully designed to prevent significant effective area losses from shadowing.

\begin{figure}
  \centering
  \includegraphics[width=1\textwidth,clip]{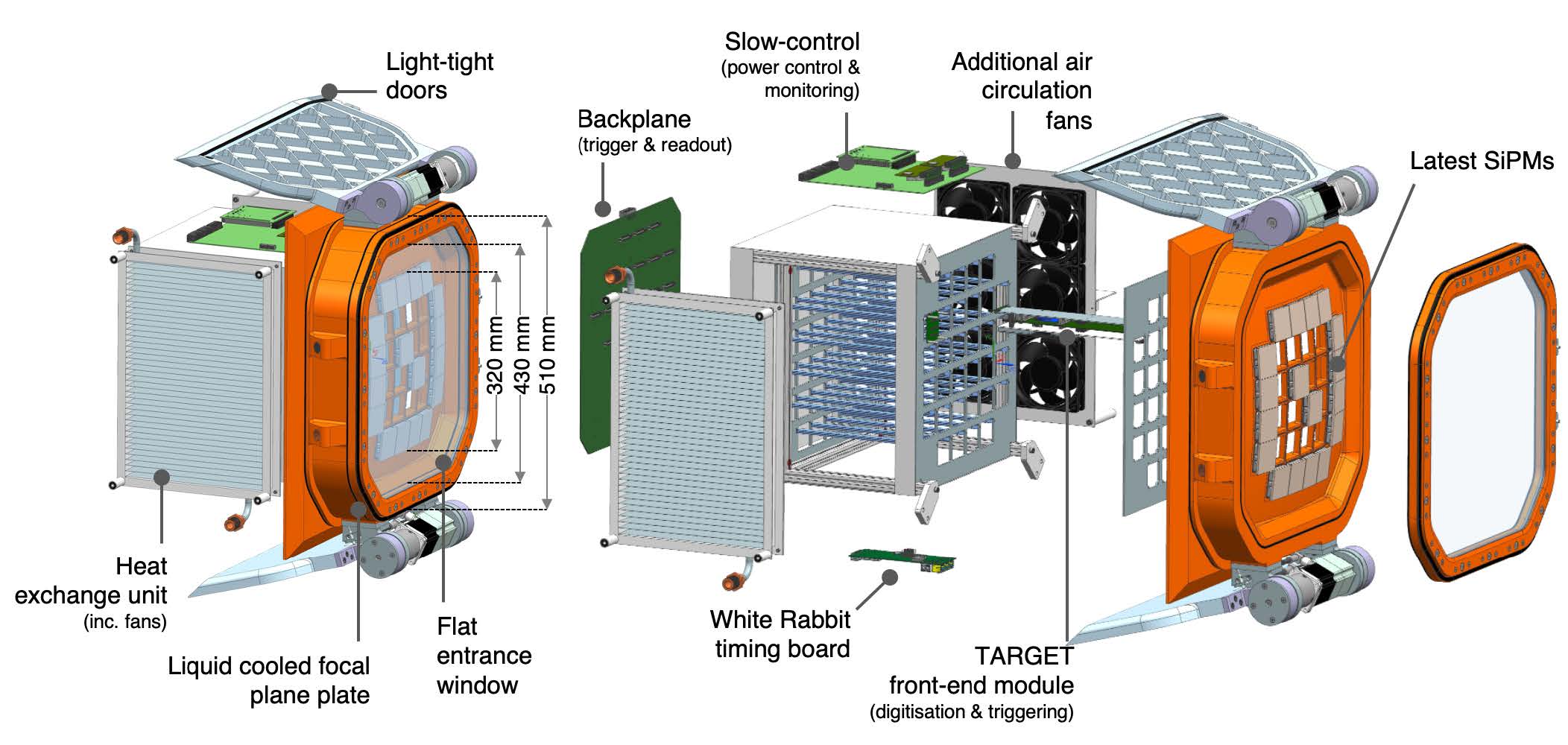}
  \caption{3D CAD mechanical model of the SST camera (2021 version). The octagonal camera window of diameter 430\,mm protects the SiPM array aligned on the focal plane. The figure is reproduced from \cite{White:2022:The-Small-Sized-Telescopes-for-the-Southern-Site-o} under the Creative Commons License (CC BY-NC-ND 4.0).}
  \label{fig:camera}
\end{figure}

\begin{figure}
  \centering
  \makebox[\textwidth][c]{
  \subfigure[]{%
    \includegraphics[width=.53\textwidth,clip]{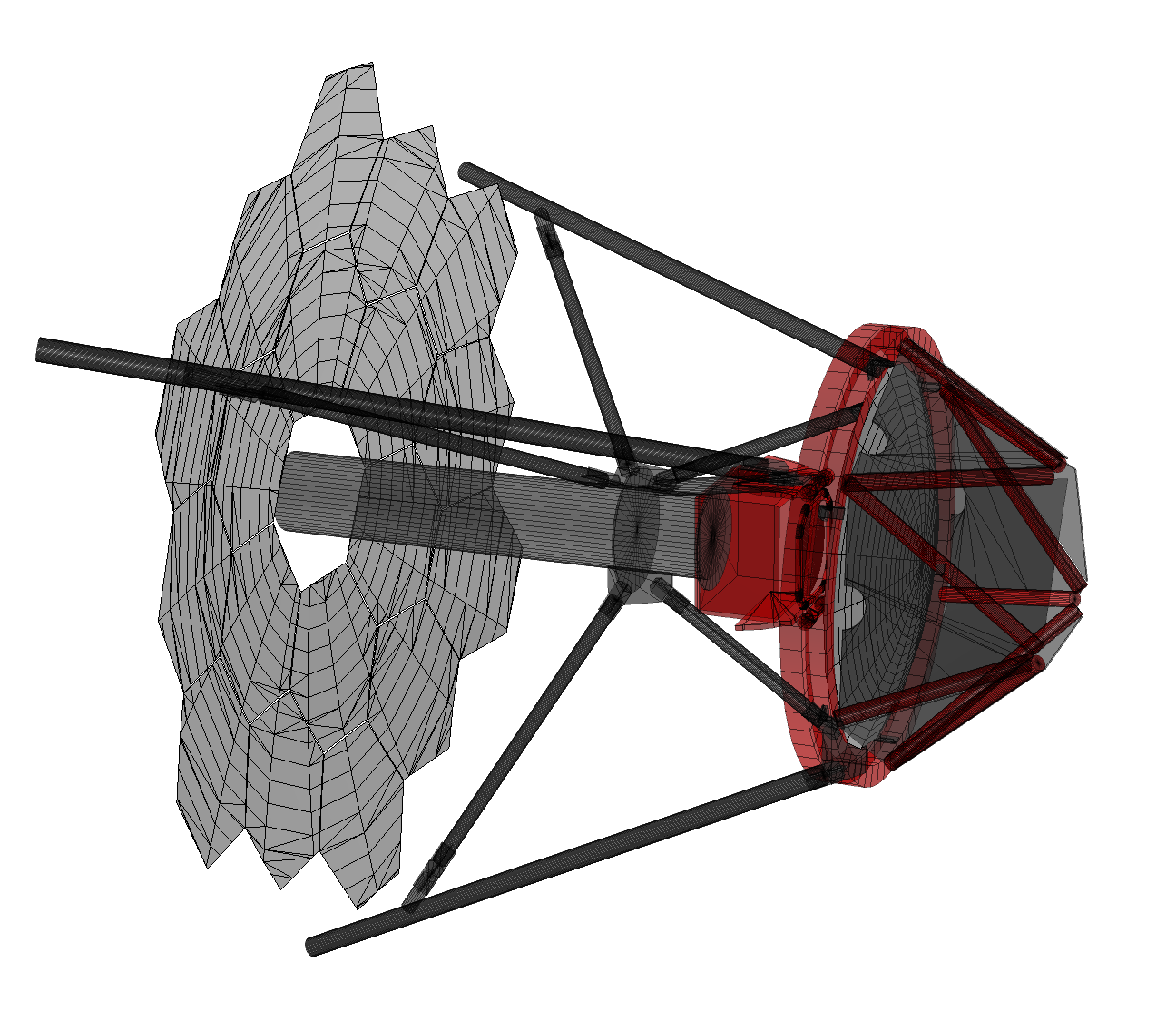}
    \label{fig:ROBAST_prod6}
  }%
  \subfigure[]{%
    \includegraphics[width=.47\textwidth,clip]{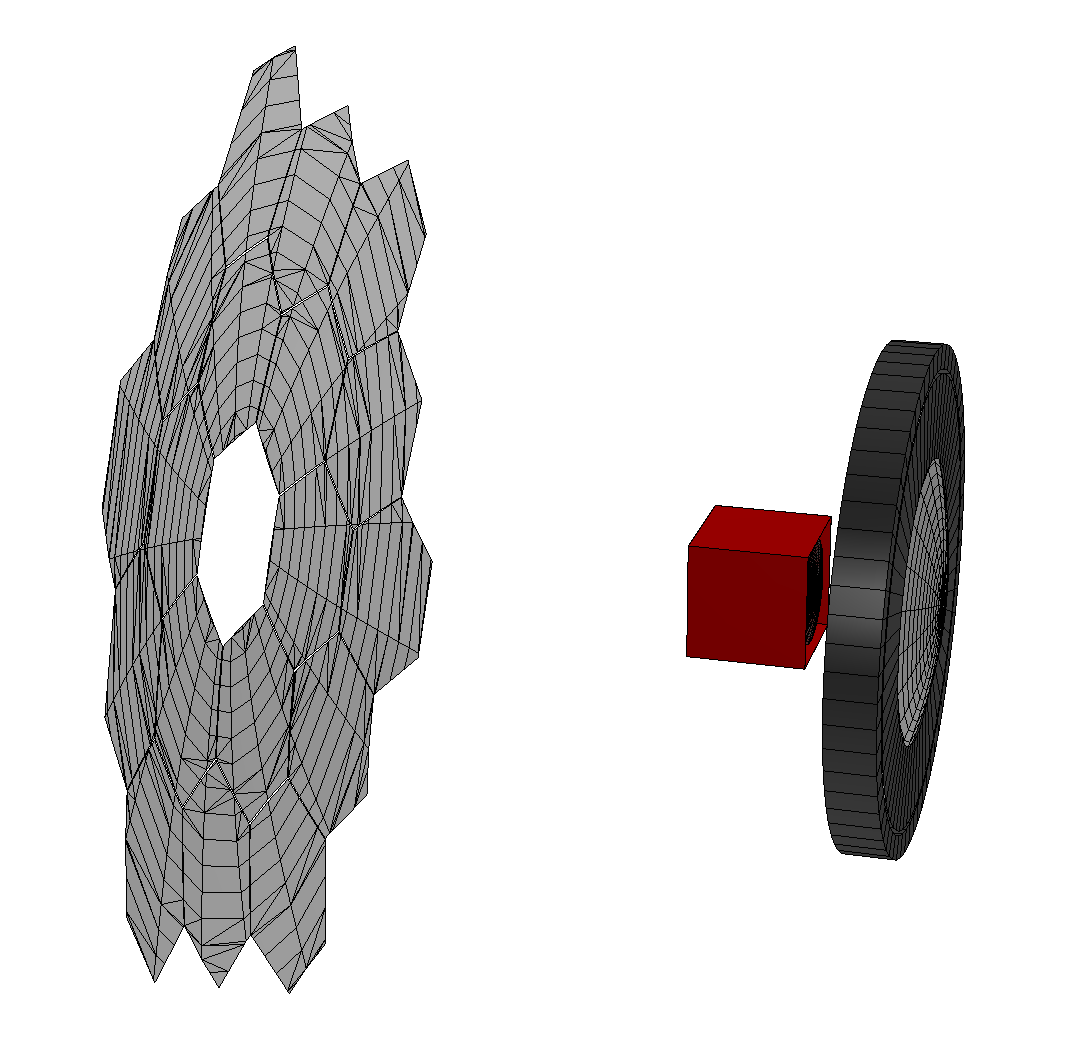}
    \label{fig:ROBAST_prod5}
  }
  }
  \caption{(a) Detailed 3D model of an SST used in full ROBAST simulations. (b) Simplified 3D model of an SST assumed in \textit{sim\_telarray}.}
\end{figure}

As shown in Fig.~\ref{fig:camera} and Fig.~\ref{fig:ROBAST_prod6}, the SST camera will have silicon photomultiplier (SiPM) tiles on the spherical focal plane to cover the $>8^\circ$ FOV with 2048 pixels. They are connected to the front-end and back-end electronics and covered by a UV-transparent flat window. If the camera doors and/or camera body enclosing the electronics are excessively large, they might interfere with the photon tracks between the primary and the secondary mirrors. However, if a small window diameter is selected to reduce the window production cost and to minimize the interference, the window edge may block the incident photons at the SiPM edges, because the angles of incidence distribute from ${\sim}30^\circ$ to ${\sim}60^\circ$. Therefore, the camera body must be sufficiently small, whereas the camera window must be sufficiently large simultaneously to consider both shadowing causes.

\begin{figure}
  \centering
  \makebox[\textwidth][c]{
  \subfigure[]{%
    \includegraphics[width=.4\textwidth,clip]{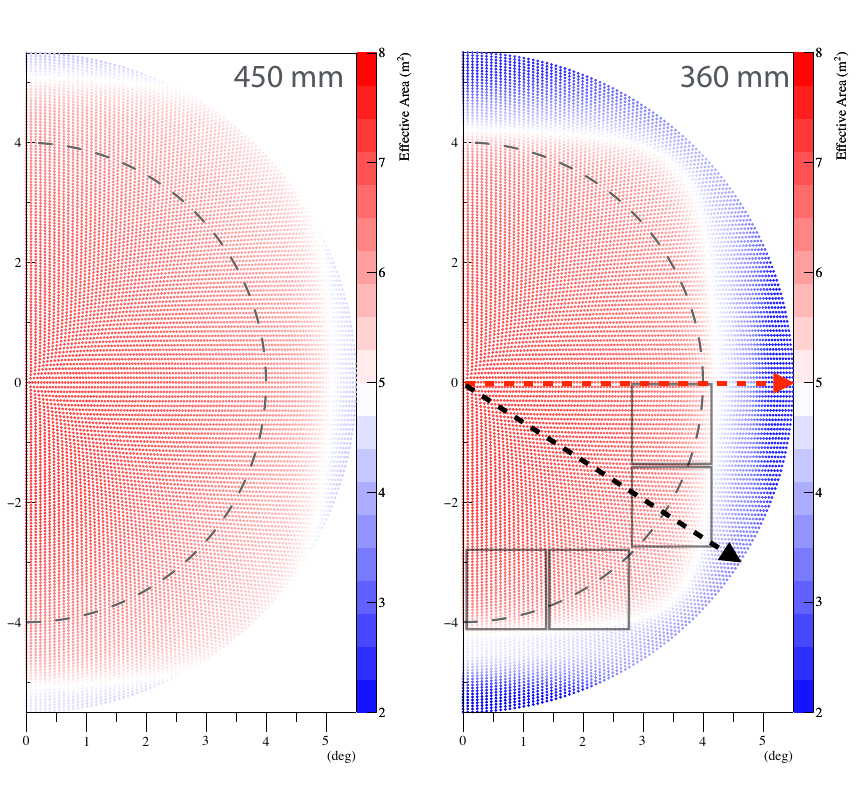}
    \label{fig:360mm_450mm}
  }%
  \subfigure[]{%
    \includegraphics[width=.6\textwidth,clip]{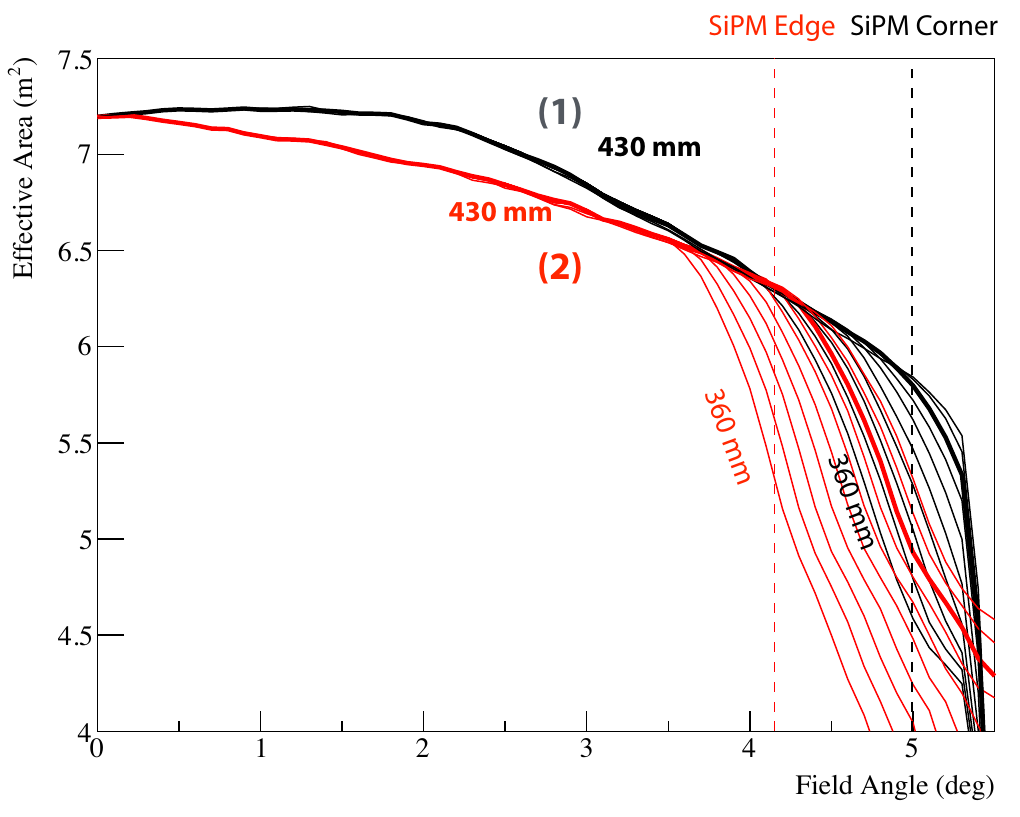}
    \label{fig:360mm_450mm_graph}
  }
  }
  \caption{(a) Distribution of the calculated effective mirror based on different camera FOV coordinates. Two window diameters are compared: 450\,mm (left) and 360\,mm (right). Four of the SiPM tile positions are represented using squares. (b) Calculated effective areas as a function of the field angle. Two slices, the center to a SiPM edge and the center to a SiPM corner, are depicted by the black and red curves (see also black and red arrows in (a)). Calculations with different window diameters (360--450\,mm in 10\,mm steps) are displayed concurrently. The curves for the 430\,mm window are highlighted with thick lines. The difference between the black and red curves in 0$^\circ$--3.5$^\circ$ is because of the asymmetric telescope masts and primary mirror.}
\end{figure}

First, we evaluated the impact of the window diameter on shadowing by simulating the SST optical system with different window diameters. The telescope structure shown in Fig.~\ref{fig:ROBAST_prod6} was fully simulated; however, the camera body structure was simplified by including only a simple camera box and an octagonal window frame. The ROBAST ray-tracing library \cite{Okumura:2022:ROBAST3,Okumura:2016:ROBAST:-Development-of-a-ROOT-based-ray-tracing-li} was used in the simulations to model the 3D structures and to trace the photon tracks. One hundred thousand parallel photons, randomly distributed in a 2.5\,m radius circle, were fallen onto the primary mirror, and the number of photons reaching the focal plane was counted to calculate the effective mirror area. This simulation was repeated for polar angles within the range $0.00$--$5.50^\circ$ ($0.05^\circ$ step) and different azimuthal angles, as shown in Fig.~\ref{fig:360mm_450mm}. The diameter of the camera in the simulation was scanned from 360 to 450\,mm with a step of 10\,mm.

As shown in Fig.~\ref{fig:360mm_450mm}, the 450\,mm camera window exhibited a smooth and moderate degradation of the effective area at large off-axis angles; however, the 360\,mm window showed a steep drop within the SiPM tile coverage. Fig.~\ref{fig:360mm_450mm_graph} shows the comparison of the changes in effective area slices along two FOV directions. Although all camera window sizes calculated in this study meet the SST requirements (${>}4^\circ$ FOV radius and ${>}5$\,m$^2$ effective area), a large nonuniformity of the effective area over the SiPM tiles is undesirable for triggering and image analysis. Therefore, a diameter of 430\,mm was selected for the camera window in our final camera design to ensure a uniform effective area.

After determining the diameter of the camera window, we evaluated the effect of the camera enclosure and window frame diameters. Increasing these parameters enables us to flexibly and easily design the internal structure and cooling system of the camera; however, an excessively large camera design may block photons passing from the primary mirror to the secondary mirror.

\begin{figure}
  \centering
  \makebox[\textwidth][c]{
  \subfigure[]{%
    \includegraphics[width=.32\textwidth,clip]{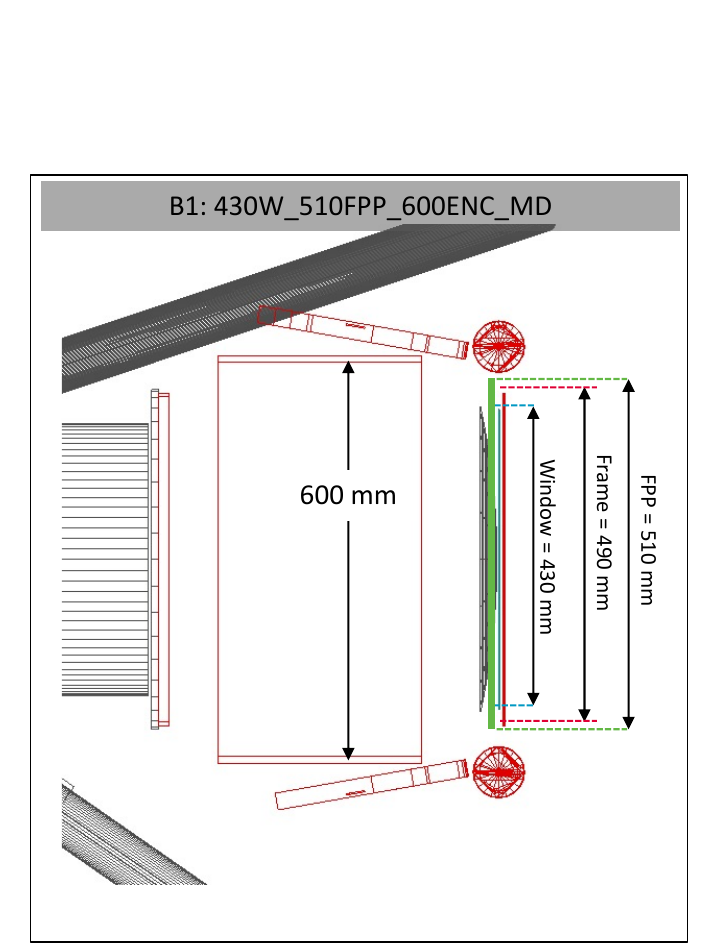}
    \label{fig:B1}
  }%
  \subfigure[]{%
    \includegraphics[width=.68\textwidth,clip]{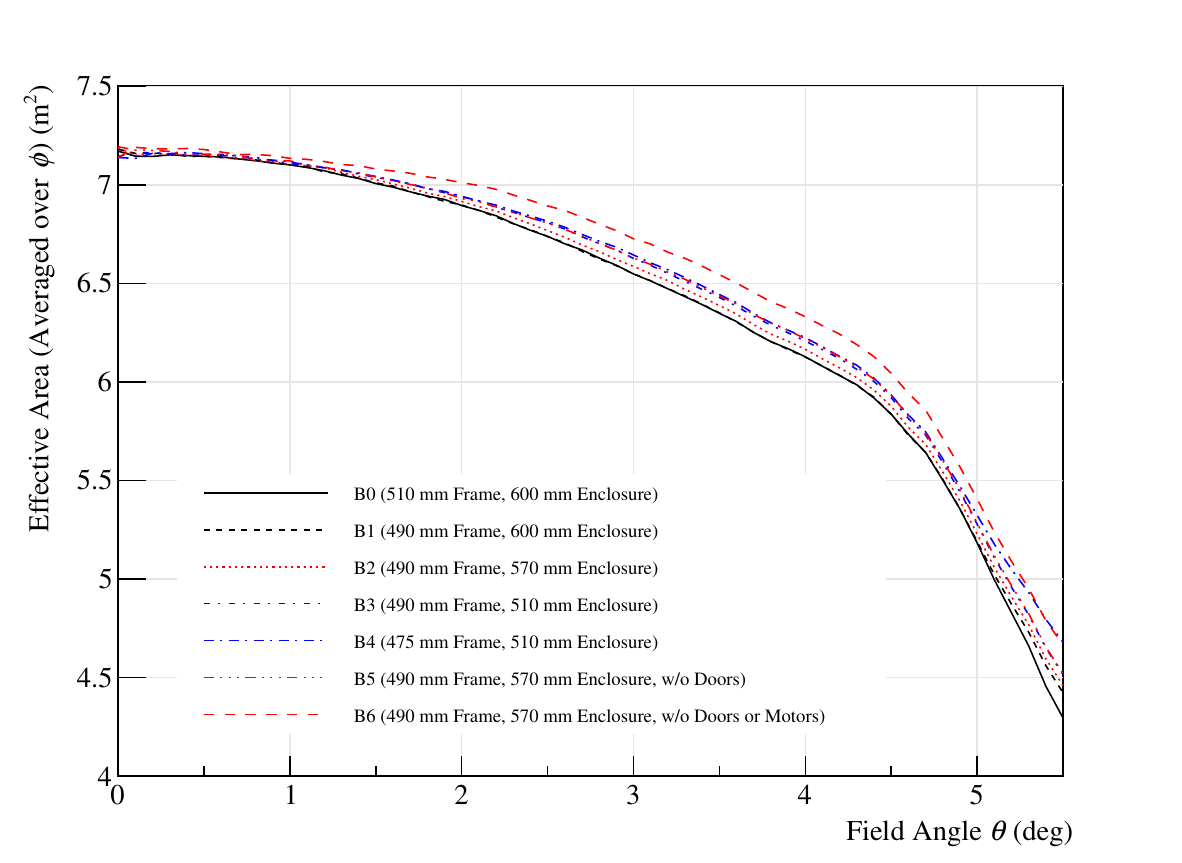}
    \label{fig:compare7}
  }
  }
  \caption{(a) Wireframe view of an SST camera design with basic dimensional parameters that are important for shadowing calculations. (b) Comparison of the effective areas of different camera dimension designs.}
\end{figure}

Fig.~\ref{fig:B1} shows a side view of one of the 3D camera models to be simulated. The effective mirror area was calculated using the same method, but with different camera enclosure and window frame sizes. Fig.~\ref{fig:compare7} compares the simulation results of seven configurations, B0--B6, including two unrealistic options, i.e., B5 and B6, without doors or motors.

The comparison of B0--B4 indicates that cameras with smaller enclosures exhibit a larger effective area as expected. However, the difference between B0 (600\,mm) and B3 (510\,mm) is only ${\sim}0.1$\,m$^2$ at a field angle of $4^\circ$. Therefore, the gain from a smaller camera design is only a few percents or less. This does not help improve the SST performance in energy bands below 1\,TeV. Therefore, from an engineering perspective, a enclosure size of 570\,mm was selected for the final camera design. A 490\,mm frame size (a 430\,nm window) was also selected because it slightly increases the effective area at around $5^\circ$ compared to a 475\,nm frame (a 415\,nm window).

\section{Final Design}
\label{sec:final_design}

As of July 2023, the SST camera design is being finalized for SST mass production \cite{Depaoli:2023:Status-of-the-SST-Camera-for-the-Cherenkov-Telesco}. The current design is based on the shadowing evaluation discussed in Section~\ref{sec:camera_design}; the design includes a 430\,mm window diameter and a 570\,mm enclosure size.

In the Cherenkov ray-tracing part of \textit{sim\_telarray}, the common MC package used in CTA \cite{Acharyya:2019:Monte-Carlo-studies-for-the-optimisation-of-the-Ch}, assumes a simple telescope and camera models as shown in Fig.~\ref{fig:ROBAST_prod5}. This is because a non-sequential ray-tracing simulation of complex telescope geometries requires significant computing power and a sequential simulation with simplified models must be employed. Instead, \textit{sim\_telarray} has a configuration parameter (function of the angles of incidence) called \textit{telescope\_transmission} that scales the effective mirror area to account for shadowing by the complex geometries.

Prior to the final camera design phase, a rather simplified telescope and camera models were assumed in the calculation of the \textit{telescope\_transmission} function. The shadowing by the structure of the minor telescope components and the camera was ignored or underestimated in the previous large MC production in CTA (``\textit{Prod5}''). Hence, we had to re-evaluate \textit{telescope\_transmission} for the latest MC production (``\textit{Prod6}'') to more accurately predict the CTA performance.

\begin{figure}
  \centering
  \includegraphics[width=.6\textwidth,clip]{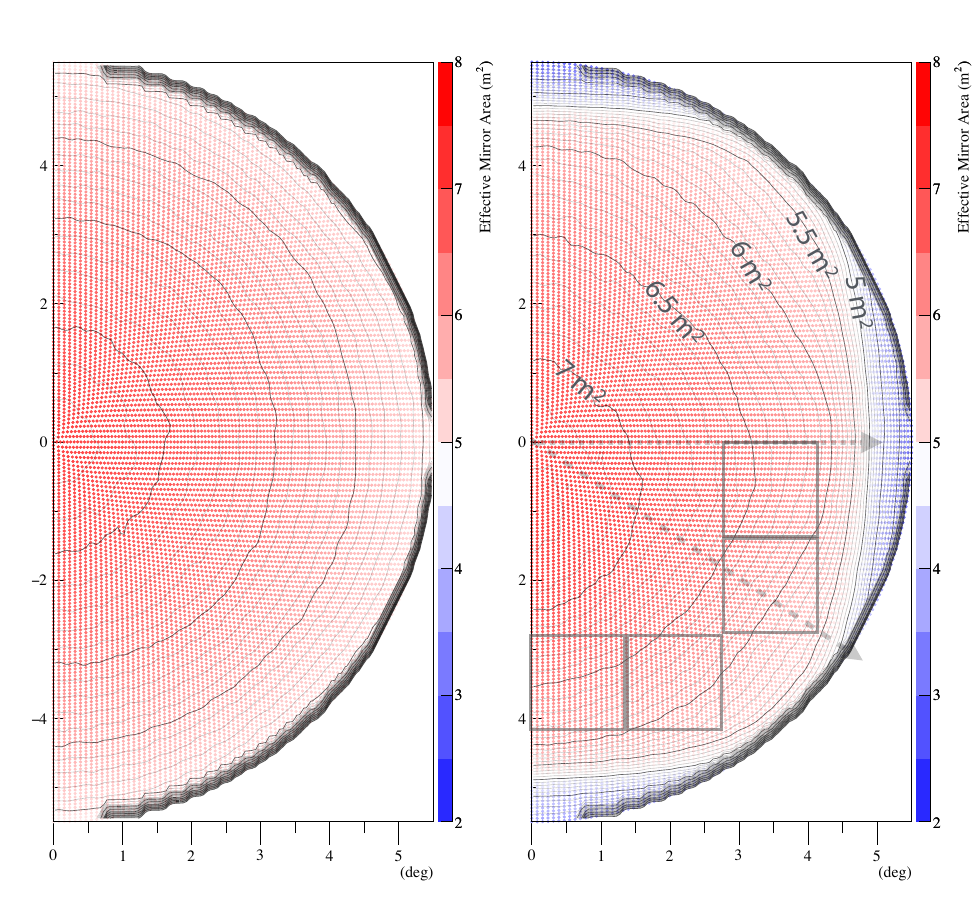}
  \caption{(Left) Distribution of the effective mirror area calculated for individual coordinates on the focal plane. Simplified 3D models of the SST telescope and camera are assumed for \textit{Prod5}. (Right) Same as left, but accurate 3D models are assumed for \textit{Prod6}. Four SiPM tile positions at the camera edges are approximately indicated by the squares.}
  \label{fig:CompareProd6_can}
\end{figure}

\begin{figure}
  \centering
  \includegraphics[width=.65\textwidth,clip]{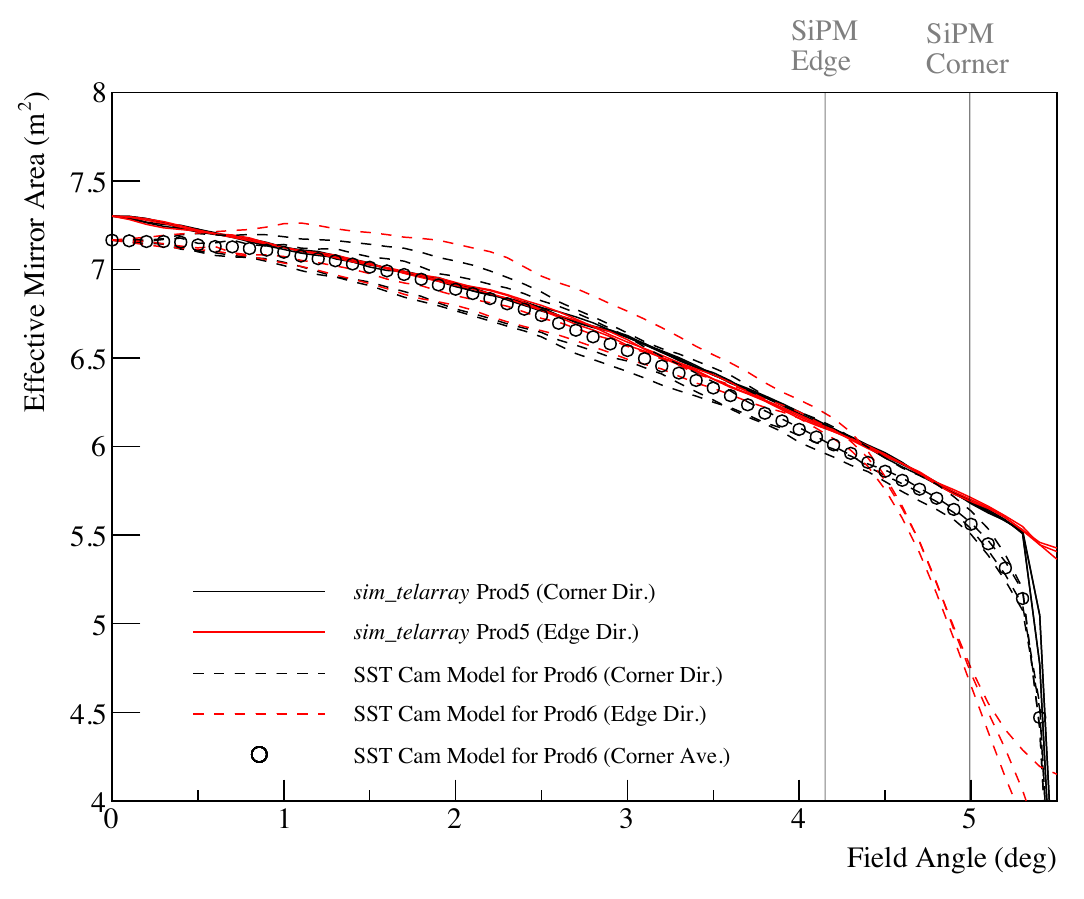}
  \caption{Comparison of the effective areas assumed in \textit{Prod5} and \textit{Prod6}. The former is axisymmetric, whereas the latter is asymmetric. Thus, several azimuthal directions exhibit small (${\sim}0.1$\,mm$^2$) variations in \textit{Prod6}.}
  \label{fig:CompareProd6_can3}
\end{figure}

Fig.~\ref{fig:CompareProd6_can} compares the ROBAST simulations for the simplified \textit{Prod5} model and a full 3D model for \textit{Prod6}. The former is axisymmetric about the optical axis (i.e., FOV center); however, the latter is asymmetric because of the full 3D implementation of the telescope masts and camera window. The nonuniformity is more visible in Fig.~\ref{fig:CompareProd6_can3}, where Fig.~\ref{fig:CompareProd6_can} is sliced in several directions. Owing to the nonuniformity in the full ROBAST simulation, the relative difference of the effective mirror area reaches approximately 10\% around $2^\circ$. Hence, the energy reconstruction of a single gamma-ray event is dependent on the direction on the camera, and the energy resolution can degrade if the nonuniformity is not considered in image analysis.

\begin{figure}
  \centering
  \includegraphics[width=.7\textwidth,clip]{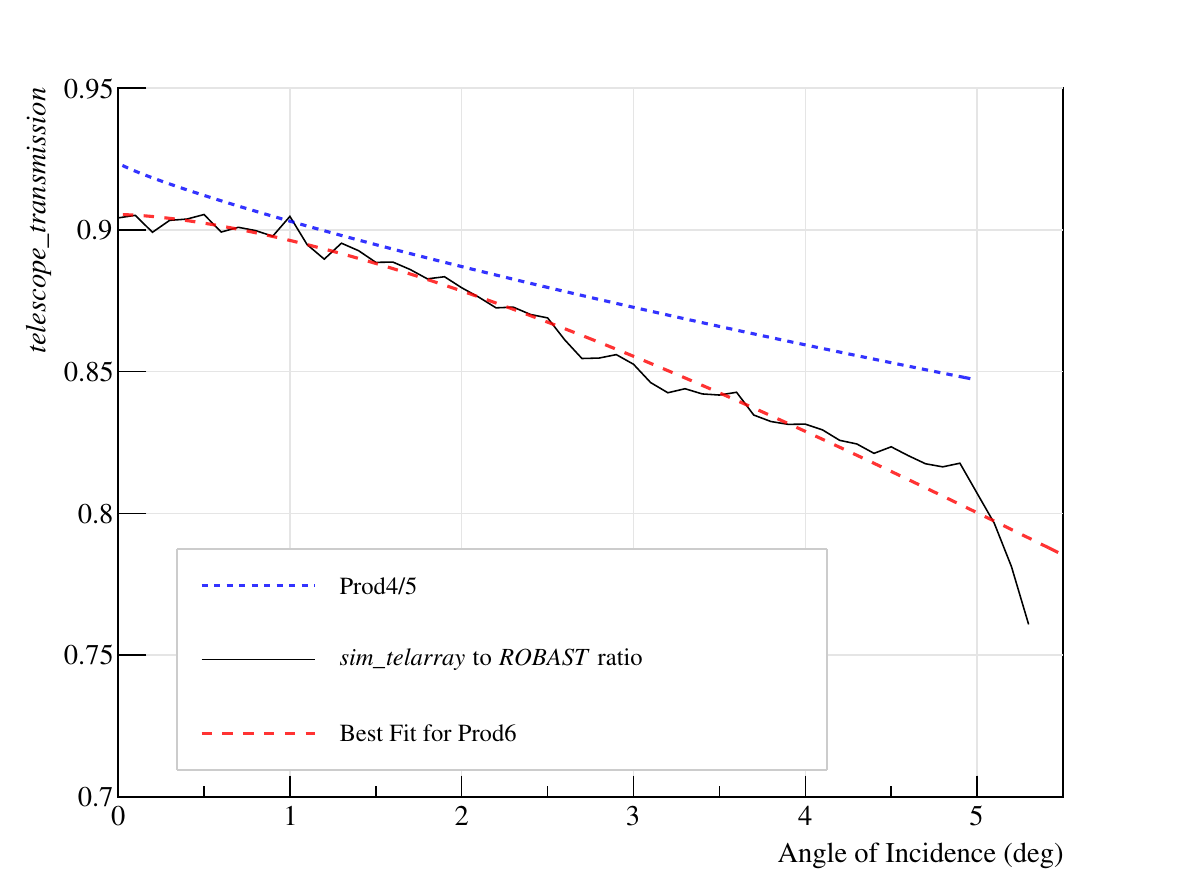}
  \caption{Comparison of the \textit{telescope\_transmission} functions used in \textit{Prod5} (blue dashed line) and \textit{Prod6} (red long-dashed line).}
  \label{fig:prod5_can_trans}
\end{figure}

In the current implementation of \textit{sim\_telarray}, the \textit{telescope\_transmission} function cannot be asymmetric. Instead, \textit{Prod6} uses a symmetric \textit{telescope\_transmission} averaged over azimuthal angles. Fig.~\ref{fig:prod5_can_trans} compares the \textit{telescope\_transmission} used in \textit{Prod5} and \textit{Prod6}. The latter was calculated by comparing the simplified \textit{sim\_telarray} simulation and full ROBAST simulation in this study. In the future version of \textit{sim\_telarray}, asymmetric \textit{telescope\_transmission} must be implemented to minimize the difference between MC simulations and real data.

\section{Conclusion}

We calculated the effective mirror area of the CTA SSTs using the ROBAST ray-tracing library and a full 3D model of the telescope and camera structure. A simulation was performed in the camera design phase to finalize a few important parameters of the camera body size. After the finalization, another simulation was performed to calculate the \textit{telescope\_transmission} parameter in \textit{sim\_telarray} for the CTA MC production \textit{Prod6}. This study revealed an asymmetric effective area over the camera FOV, which must be considered in the Cherenkov image analysis for future gamma-ray observations.

\providecommand{\href}[2]{#2}\begingroup\raggedright\endgroup

\acknowledgments

This study was supported by JSPS KAKENHI Grant Numbers JP18KK0384, JP20H01916, and JP23H04897.

%
%
%


\begin{thebibliography}{1}

\bibitem{Vassiliev:2007:Wide-field-aplanatic-two-mirror-telescopes-for-gro}
V.~Vassiliev, S.~Fegan and P.~Brousseau, \emph{Wide field aplanatic two-mirror
  telescopes for ground-based $\gamma$-ray astronomy},
  \href{https://doi.org/10.1016/j.astropartphys.2007.04.002}{\emph{Astropart.~Phys.}
  {\bfseries 28} (2007) 10}.

\bibitem{Adams:2021:Detection-of-the-Crab-Nebula-with-the-9.7-m-protot}
C.~Adams, R.~Alfaro, G.~Ambrosi, M.~Ambrosio, C.~Aramo, T.~Arlen et~al.,
  \emph{Detection of the {Crab Nebula} with the 9.7 m prototype
  {Schwarzschild--Couder} telescope},
  \href{https://doi.org/https://doi.org/10.1016/j.astropartphys.2021.102562}{\emph{Astroparticle
  Physics} {\bfseries 128} (2021) 102562}.

\bibitem{Lombardi:2020:First-detection-of-the-Crab-Nebula-at-TeV-energies}
S.~Lombardi, {Catalano, O.}, {Scuderi, S.}, {Antonelli, L. A.}, {Pareschi, G.},
  {Antolini, E.} et~al., \emph{First detection of the {Crab Nebula} at {TeV}
  energies with a {Cherenkov} telescope in a dual-mirror {Schwarzschild-Couder}
  configuration: the {ASTRI-Horn} telescope},
  \href{https://doi.org/10.1051/0004-6361/201936791}{\emph{Astron.~Astrophys.}
  (2020) A22}.

\bibitem{White:2022:The-Small-Sized-Telescopes-for-the-Southern-Site-o}
R.~{White}, J.P.~{Amans}, D.~{Berge}, G.~{Bonanno}, R.B.~{Bose}, A.M.~{Brown}
  et~al., \emph{{The Small-Sized Telescopes for the Southern Site of the
  Cherenkov Telescope Array}},  in \emph{37th International Cosmic Ray
  Conference}, p.~728, Mar., 2022,
  \href{https://doi.org/10.22323/1.395.0728}{DOI}
  [\href{https://arxiv.org/abs/2110.14527}{{\ttfamily 2110.14527}}].

\bibitem{Okumura:2022:ROBAST3}
A.~Okumura, \emph{{ROBAST}~3},  in \emph{Proc.~37th~Int.~Cosmic~Ray~Conf.},
  vol.~395, p.~183, 2022, \href{https://doi.org/10.22323/1.395.0183}{DOI}.

\bibitem{Okumura:2016:ROBAST:-Development-of-a-ROOT-based-ray-tracing-li}
A.~Okumura, K.~Noda and C.~Rulten, \emph{{ROBAST}: Development of a
  {ROOT}-based ray-tracing library for cosmic-ray telescopes and its
  applications in the {Cherenkov Telescope Array}},
  \href{https://doi.org/10.1016/j.astropartphys.2015.12.003}{\emph{Astropart.~Phys.}
  {\bfseries 76} (2016) 38}.

\bibitem{Depaoli:2023:Status-of-the-SST-Camera-for-the-Cherenkov-Telesco}
D.~Depaoli, \emph{Status of the {SST} camera for the {Cherenkov Telescope
  Array}},  in \emph{Proc.~38th~Int.~Cosmic~Ray~Conf.}, p.~771, 2023,
  \href{https://doi.org/https://doi.org/10.22323/1.444.0771}{DOI}.

\bibitem{Acharyya:2019:Monte-Carlo-studies-for-the-optimisation-of-the-Ch}
A.~Acharyya, I.~Agudo, E.~Ang{\"u}ner, R.~Alfaro, J.~Alfaro, C.~Alispach
  et~al., \emph{{Monte Carlo} studies for the optimisation of the {Cherenkov
  Telescope Array} layout},
  \href{https://doi.org/https://doi.org/10.1016/j.astropartphys.2019.04.001}{\emph{Astroparticle
  Physics} {\bfseries 111} (2019) 35}.

\end{thebibliography}
\end{document}